\documentclass[prb,showpacs,twocolumn,superscriptaddress,aps,a4paper]{revtex4}
\usepackage{pstricks,pst-node,pst-text,pst-3d,graphpap,pst-plot}
\usepackage{dcolumn}\usepackage{verbatim}
\usepackage{amsmath}
\usepackage{graphicx}
\usepackage{latexsym}
\usepackage{amsfonts}
\usepackage{amssymb}
\DeclareGraphicsExtensions{.pdf,.gif,.jpg}

\newcommand{\be}{\begin{equation}}
\newcommand{\ee}{\end{equation}}
\newcommand{\beq}{\begin{eqnarray}}
\newcommand{\eeq}{\end{eqnarray}}

\begin{document}

\def\bbe{\mbox{\boldmath $e$}}
\def\bbf{\mbox{\boldmath $f$}}
\def\bg{\mbox{\boldmath $g$}}
\def\bh{\mbox{\boldmath $h$}}
\def\bj{\mbox{\boldmath $j$}}
\def\bq{\mbox{\boldmath $q$}}
\def\bp{\mbox{\boldmath $p$}}
\def\br{\mbox{\boldmath $r$}}
\def\bz{\mbox{\boldmath $z$}}

\def\bfzero{\mbox{\boldmath $0$}}
\def\bfone{\mbox{\boldmath $1$}}

\def\dr{{\rm d}}

\def\tb{\bar{t}}
\def\zb{\bar{z}}

\def\tgb{\bar{\tau}}

\def\bC{\mbox{\boldmath $C$}}
\def\bG{\mbox{\boldmath $G$}}
\def\bH{\mbox{\boldmath $H$}}
\def\bK{\mbox{\boldmath $K$}}
\def\bM{\mbox{\boldmath $M$}}
\def\bN{\mbox{\boldmath $N$}}
\def\bO{\mbox{\boldmath $O$}}
\def\bQ{\mbox{\boldmath $Q$}}
\def\bR{\mbox{\boldmath $R$}}
\def\bS{\mbox{\boldmath $S$}}
\def\bT{\mbox{\boldmath $T$}}
\def\bU{\mbox{\boldmath $U$}}
\def\bV{\mbox{\boldmath $V$}}
\def\bZ{\mbox{\boldmath $Z$}}

\def\bcalS{\mbox{\boldmath $\mathcal{S}$}}
\def\bcalG{\mbox{\boldmath $\mathcal{G}$}}
\def\bcalE{\mbox{\boldmath $\mathcal{E}$}}

\def\bgG{\mbox{\boldmath $\Gamma$}}
\def\bgL{\mbox{\boldmath $\Lambda$}}
\def\bgS{\mbox{\boldmath $\Sigma$}}

\def\bgr{\mbox{\boldmath $\rho$}}
\def\bgs{\mbox{\boldmath $\sigma$}}

\def\a{\alpha}
\def\b{\beta}
\def\g{\gamma}
\def\G{\Gamma}
\def\d{\delta}
\def\D{\Delta}
\def\e{\epsilon}
\def\ve{\varepsilon}
\def\z{\zeta}
\def\h{\eta}
\def\th{\theta}
\def\k{\kappa}
\def\l{\lambda}
\def\L{\Lambda}
\def\m{\mu}
\def\n{\nu}
\def\x{\xi}
\def\X{\Xi}
\def\p{\pi}
\def\P{\Pi}
\def\r{\rho}
\def\s{\sigma}
\def\S{\Sigma}
\def\t{\tau}
\def\f{\phi}
\def\vf{\varphi}
\def\F{\Phi}
\def\c{\chi}
\def\w{\omega}
\def\W{\Omega}
\def\Q{\Psi}
\def\q{\psi}

\def\ua{\uparrow}
\def\da{\downarrow}
\def\de{\partial}
\def\inf{\infty}
\def\ra{\rightarrow}
\def\bra{\langle}
\def\ket{\rangle}
\def\grad{\mbox{\boldmath $\nabla$}}
\def\Tr{{\rm Tr}}
\def\Re{{\rm Re}}
\def\Im{{\rm Im}}
\def\hc{{\rm h.c.}}

\title{Bouncing transient currents and SQUID-like voltage in nano devices at half filling}

\author{Michele Cini}
\affiliation{Dipartimento di Fisica, Universit\`a di Roma Tor
Vergata, Via della Ricerca Scientifica 1, 00133 Rome, Italy}
\affiliation{Istituto Nazionale
di Fisica Nucleare, Laboratori Nazionali di Frascati, Via E. Fermi 40, 00044 Frascati, Italy}

\author{Enrico Perfetto}
\affiliation{Unit\`a CNISM, Universit\`a di Roma Tor Vergata,
Via della Ricerca Scientifica 1, 00133 Rome, Italy}

\author{Chiara Ciccarelli}
\affiliation{Dipartimento di Fisica, Universit\`a di Roma Tor
Vergata, Via della Ricerca Scientifica 1, 00133 Rome, Italy}

\author{Gianluca Stefanucci}
\affiliation{Dipartimento di Fisica, Universit\`a di Roma Tor
Vergata, Via della Ricerca Scientifica 1, 00133 Rome, Italy}
\affiliation{Istituto Nazionale
di Fisica Nucleare, Laboratori Nazionali di Frascati, Via E. Fermi 40, 00044 Frascati, Italy}
\affiliation{European Theoretical Spectroscopy Facility (ETSF)}

\author{Stefano Bellucci}
\affiliation{Istituto Nazionale
di Fisica Nucleare, Laboratori Nazionali di Frascati, Via E. Fermi 40, 00044 Frascati, Italy}

\begin{abstract}
Nanorings  asymmetrically connected to wires show  different kinds
of quantum interference  phenomena under sudden excitations and in
steady current conditions.  Here we contrast the transient current
caused by an abrupt  bias to  the magnetic effects at constant
current. A repulsive impurity  can cause  charge build-up in one of
the arms and reverse current spikes.
 Moreover, it can
cause transitions from laminar current flow to vortices, and also
change the chirality of the vortex. The magnetic behavior of these
devices is also very peculiar.
Those nano-circuits which consist of an odd number of atoms behave in a fundamentally different manner compared to those which consist of an even number of atoms.
The circuits  having an odd number of
sites  connected to long enough symmetric wires   are diamagnetic; they  display half-fluxon periodicity
induced by many-body symmetry  even in the absence of
electron-phonon and electron-electron interactions.    In principle
one can  operate a new kind of quantum interference device without
superconductors. Since there is no gap and  no  critical temperature,  one predicts
qualitatively the same behavior at and above room temperature,
although with a reduced current. The circuits with even site numbers, on the other hand, are paramagnetic.

\end{abstract}

\pacs{72.10.Bg,85.25.Dq,74.50.+r}


\maketitle

\section{Introduction}
The last decade has witnessed a growing interest in the persistent
currents in quantum rings threaded by a magnetic flux $\phi.$ This
problem has many variants.  The rings of interest may contain  impurities, may  interact with quantum dots or reservoirs. Moreover,  both  continuous and discrete formulations
have been used to date, with similar results. Aligia\cite{aligia}
modeled the persistent currents in a ring with an embedded quantum
dot. Theoretical approaches to one-dimensional and
quasi-one-dimensional quantum rings with a few electrons are
reviewed by S.Viefers and coworkers\cite{viefers}; see also the
recent review by S. Maiti\cite{maiti} where it is pointed out that
central issues like the diamagnetic or the paramagnetic sign of the
low-field currents of isolated rings are still unsettled. The present paper, instead, is
 devoted  to rings connected  to circuits.  It is clear that the connection to wires substantially modifies the  problem, but here we point out that some of the modifications are not obvious and lead   to quite interesting and novel consequences.

 Simple quantum rings can be connected to
biased wires in such a way that the current flows through
inequivalent paths. In the present, exploratory paper, we consider a
tight-binding ring with $N$ sites attached to two one-dimensional
leads and  specialize on the half-filled system. We show that this
innocent-looking situation produces peculiar phenomena like {\em
bouncing transients} i.e. current spikes in the reverse direction,
and a new sort of simulated pairing. These systems could find
applications in spintronic or fast electronic devices exploiting
spikes in the onset currents, or also in steady current conditions
when used to measure local magnetic fields.  Thus, our motivation is
twofold. On one hand, we look  for the conditions that can produce
novel phenomena in the  transient current when the system is biased.
On the other hand we are interested  in the bias that develops
across the system in steady current conditions when it is used like
a SQUID. After presenting the model in the next Section, in Section
\ref{Formalism} we present the formalism and in Section
\ref{Switch-on} apply it to situations where transient currents
through the circuit are large and appear to bounce to a direction
contrary to the main stream. We then discuss the currents inside the
device, with vortices that can be clockwise or counterclockwise
depending on the parameters. The nontrivial topology allows interference effects of the Aharonov-Bohm type in one-body experiments, but  in addition here, a many-body symmetry comes into play when the ring has an odd number of atoms. The interplay of symmetry and topology
may lead to a  diamagnetic behavior with a half-fluxon periodicity   which looks like a typical
superconducting pattern as shown in Section \ref{doubleminimum}.  The possible
operation of the circuit as a magnetometer is also suggested. Our
main results are summarized in the concluding Section
\ref{conclusions}.

\section{Model}\label{model}

\begin{figure}[htbp]
\includegraphics*[width=.47\textwidth]{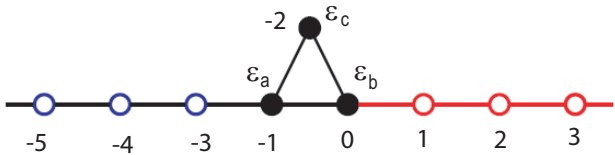}
\caption{Sketch of the site numbering and site energies of the
triangular ring, shown  with part of the leads. The impurity is
denoted by its energy $\e_{c}$ while we shall set $\e_{a}=\e_{b}=0.$
The hopping integrals in the absence of magnetic field  are
$t_{ab}=t_{bc}=t_{ca}=t_{dev}$.} \label{triangolo}
\end{figure}
The Hamiltonian describing the left ($L$) and right ($R$)
one-dimensional leads is \be\label{leadsh}
H_{\a}=t_{h}\sum_{m=0}^{\inf}( c^{\dag}_{m,\a}c_{m+1,\a}+ {\rm
h.c.}), \ee where  $t_{h}$ is the hopping integral between nearest
neighbor sites and  $c_{m+1,\a}$ annihilates an electron at site
$m+1$ in wire $\a=L,R$. The spin indices  are not shown in order to
simplify the notation. The energy window of both $L$ and $R$
continua is $(-2|t_{h}|,2|t_{h}|)$ and the half-filled system
corresponds to having a chemical potential $\m_{\rm}=0$. In this
work we consider $N$-sided  polygonal devices described by the
Hamiltonian \be H_{\rm ring}=\sum_{<m,n>=1}^{N} (t_{mn}
d^{\dag}_{m}d_{n}+\hc) \label{ringham} \ee where $d_{n}$ are Fermion
annihilation operators inside the device and $t_{mn}$ is the hopping
between site $m$ and site $n$. In the rest of the paper we
specialize to the case $t_{mn}=0$ if $m$ and $n$ are not nearest
neighbors and $t_{mn} \equiv t_{dev}e^{i\a(m,n)}$, where
$e^{i\a(m,n)}$ is a phase factor, otherwise. The simplest example is
the triangular model ($N=3$) shown in Figure 1. In this case  we
also label the sites with letters and consider   a non-magnetic
impurity at the vertex site $c$ of strength $\e_{c}$: \be H_{\rm
ring}=\sum_{<m,n>\in \{a,b,c\}} (t_{mn}
d^{\dag}_{m}d_{n}+\hc)+\e_{c}n_{c}, \label{ringham2} \ee with
$n_{c}=d^{\dag}_{c}d_{c}$.

The wires will be attached at a couple of sites  in the ring that
will be specified case by case  below by using a tunneling
Hamiltonian $H_{T}$ with hopping parameter $t_{h}$. Thus, the
equilibrium Hamiltonian  reads \be H_0=H_{L}+H_{R}+H_{\rm
ring}+H_{T} \label{acca0} \ee while for times $t>0$ \be
H(t)=H_{0}+H_{\rm bias} (t), \ee where \be H_{\rm
bias}(t)=\sum_{\a=L,R}\sum_{m=0}^{\inf}V_{\a}(t)n_{m,\a}, \ee
with $n_{m\a}=c^{\dag}_{m\a}c_{m\a}$. The bias
shifts the energy of the $\a$ sites by a constant amount, with a time dependence
$V_{\a}(t)$.

The  electron number current operator  between sites $m$ and $n$ connected by a bond
with hopping integral $\t_{mn}$ ($\t_{mn}=t_{h}$ or $t_{dev}e^{i
\a(m,n)}$ inside the device) is determined\cite{caroli} by imposing
the continuity equation \be J_{mn}=-\frac{i}{\hbar}
\t_{mn}c^{\dag}_{n}c_{m}+{\rm h.c.}.
\label{scossa} \ee  For ring sites, $c$ operators will be replaced by
$d$ ones.

\section{Formalism}\label{Formalism}
\label{time} In the partition-free approach\cite{cini80} the formula
for the time-dependent averaged current through the $m-n$ bond reads
\be \langle J_{mn}(t)\rangle =\Tr\left[
\hat{f}^{(0)}U^{\dag}(t)J_{mn}U(t) \right], \ee with \be
\hat{f}^{(0)}=\frac{1}{e^{\b(H_{0}-\m)}+1}, \ee the Fermi function
computed at the equilibrium Hamiltonian $H_{0}$ and $U(t)$ the
evolution operator. In the actual calculations we have adopted a
local view and write the electron number
current \beq \label{local} \langle J_{mn}(t)\rangle
=\nonumber\\
\frac{2e}{\hbar}\Im \left [\t_{mn} \sum_{rs}\langle n |
U^{\dagger}(t) |r\rangle\langle s|U(t) |m\rangle f(r,s)\right ], \eeq
where $f(r,s)=\langle r|\hat{f}^{(0)}| s\rangle$ is the matrix
element of $\hat{f}^{(0)}$ between one-particles states localized at
sites $r$ and $s$. In this way we observed that far sites in the
wires come into play one after another with a clear-cut  delay.   One can simulate infinite leads  with
wires consisting of $ l$ sites and obtain quite accurate currents for
times up to $t=\frac{ \hbar}{t_{h}}{l\over 3};$ the absence of more distant sites does not change the results in any appreciable measure. Eventually, when $t$ is increased at fixed length $l$, the information that the wires are finite arrives quite suddenly and a fast
drop of the current takes place.\cite{psc.2008}

Another useful expression for the bond-current which holds for step-function switching of the bias can be obtained by inserting into Equation (\ref{local}) complete sets $\psi_{1},\psi_{2}$ of $H$ eigenstates (in the presence of the bias) with energy eigenvalues $E(\psi_{1}), E(\psi_{2})$; one gets

\beq\label{oscilla}
\langle J_{mn}(t)\rangle =-\frac{2e}{\hbar}Im \left (t_{mn}\sum_{\psi_{1},\psi_{2}}
e^{i \frac{(E(\psi_{1})-E(\psi_{2})t}{\hbar}}\right.\nonumber\\
\left.\langle n |\psi_{1}\rangle \langle\psi_{2}|m\rangle \sum_{n_{1},n_{2}}\langle \psi_{1}|n_{1}\rangle\langle n_{2}|\psi_{2}\rangle f(n_{1},n_{2})\right).\eeq
This informs us about the frequency spectrum of the current response.
The frequencies  arise from energy differences between the eigenstates
of the Hamiltonian of the full circuit with the bias included.
The  weights at a given bond  depend in a simple way on  the
eigenfunctions at  the bond and on the equilibrium occupation of
single-electron states. If there are sharp discrete states outside
the continuum, $\langle J_{mn}(t)\rangle$ has an oscillatory component, otherwise it tends\cite{cini80} to the current-voltage characteristics asymptotically as $t\rightarrow\infty$.

\section{Switch-on currents}
\label{Switch-on}

We study the triangular model of Figure 1 with
$t_{ab}=t_{bc}=t_{ca}=t_{dev}$  and  $t_{dev}=t_{h}$ for the sake of
definiteness. Our main idea in the model calculations has been to
study the $\e_{c}$ dependence of the transient in order to look for
marked  out-of-equilibrium behavior, like vortex formation, or large
charge build-up followed by strong current spikes.   Also, one is
interested in conditions that produce a fast change of the response
with $\e_{c},$  since possibly in such situations magnetic
impurities can produce strong spin polarization.

Below we present numerical results for 150 sites in the leads (which
guarantee an accurate propagation up to times $20\hbar/t_{h}$) and
switch on a constant bias $V_{L}=V=0.5$ and $V_{R}=0$ at $t=0$. In
the steady state it tends to produce a negative  local current flow
(in the sense that the electron number  current goes from the right
to the left wire).

\subsection{Total current}
In Figure 2 we show the time-dependence of the number current in
$\frac{t_{h}}{\hbar}$ units on the   bond between sites -3 and -4
for different values of $\epsilon_{c}.$

\subsubsection{Long-time limit}

The times we are considering are short enough to enable us to use
the finite-lead scheme specified above, however the system already
clearly approaches the steady state for all $\e_{c}$. The effect of
the impurity at site $c$ on the asymptotic current is much stronger
than one could have expected from any classical analogue. The
current at $\e_{c}=-1$ (in units of $t_{h}$) is more than an order
of magnitude larger than it is at $\e_{c}=+1.$  Moreover, in a range
around $\e_{c}=0$ the negative current increases (that is, its
absolute value decreases) with increasing $\e_{c},$ as one could
expect if the repulsive site were an obstruction to the current
flow. However this view is not in line with the fact that a further
increase of   $\e_{c}$  increases the conductivity of the device.
The correct interpretation is that the conductivity is ruled by
quantum interference between the $a-b$ and $a-c-b$ paths,  particularly
by electrons around the Fermi-level.

\subsubsection{Short and intermediate times}

Here we are in position to see how this quantum interference
develops in time. It  takes  a time $\sim \frac{ \hbar}{t_{h}}$ for
the current to go from zero to the final order-of-magnitude;  then
for some $\frac{\hbar}{t_{h}}$ large oscillations occur, which
appear to be damped with a characteristic time $\sim 10
\frac{\hbar}{t_{h}}.$
The oscillations have characteristic
frequencies that  in accordance with Equation  (\ref{oscilla})  increase by increasing  the hopping matrix
elements  or  modifying the device in any way that enhances the
energy level differences. It can be seen that the damping of the
oscillations is not exponential, and  characteristic times of the
order of 10 $\frac{t_{h}}{\hbar}$ are also noticeable.

\begin{figure}[htbp]
\includegraphics*[width=.47\textwidth]{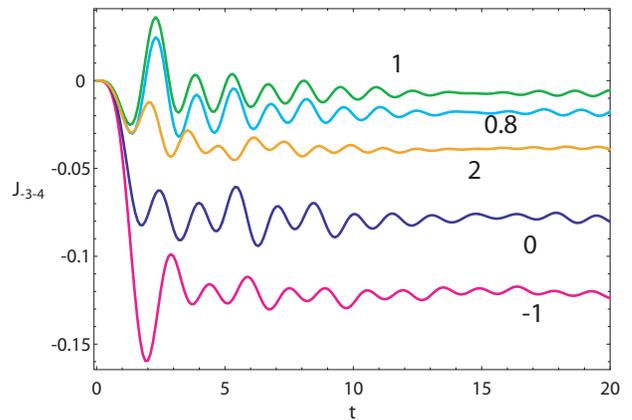}
\caption{Time-dependence of the current on the   bond between the sites
-3 and -4. The numbers shown close to each curve indicate  the
values of $\epsilon_{c}$. From bottom $\e_{c}=-1,0,2,08,1$ in units of $t_{h}$. Here and
below $t$ is in  $\frac{\hbar}{t_{h}}$ units and the current is in units of $\frac{t_{h}}{\hbar}$. }
\label{vort0}
\end{figure}

Figure \ref{vort0} shows  that at $\e_{c}=0$ the current quickly
approaches $-0.08 \frac{t_{h}}{\hbar}$, but at short times  the
spread of values of the currents is larger than it is at asymptotic
times. A negative impurity like  $\e_{c}=-1$ produces  a
spike whose magnitude exceeds by $\sim 30$
per cent the steady state value
$\sim -0.12 \frac{t_{h}}{\hbar}$, while a positive $\e_{c}$ reduces
the conductivity of the device. The  dependence of the current on
$\e_{c}$ and time is involved, with the $\epsilon_{c}=1$ case which
does not belong to the region delimited by $\epsilon_{c}=0$ and
$\epsilon_{c}=2$ curves. The $\e_{c}=0.8$ and $\e_{c}=1.0$ curves
that  produce small negative currents at long times are most
interesting. They even  go positive for some time interval $\sim
\frac{\hbar}{t_{h}}$. Positive currents go backwards. This bouncing
current is  a  quantum interference  effect, which takes the system
temporarily but  dramatically out of equilibrium and in
counter-trend to the steady state. The main spike of reversed
current is much larger than the long-time direct current  response.
Next, to understand what produces the bouncing current,  we look at
the transport inside the triangular device.

\begin{figure}[htbp]
\includegraphics*[width=.47\textwidth]{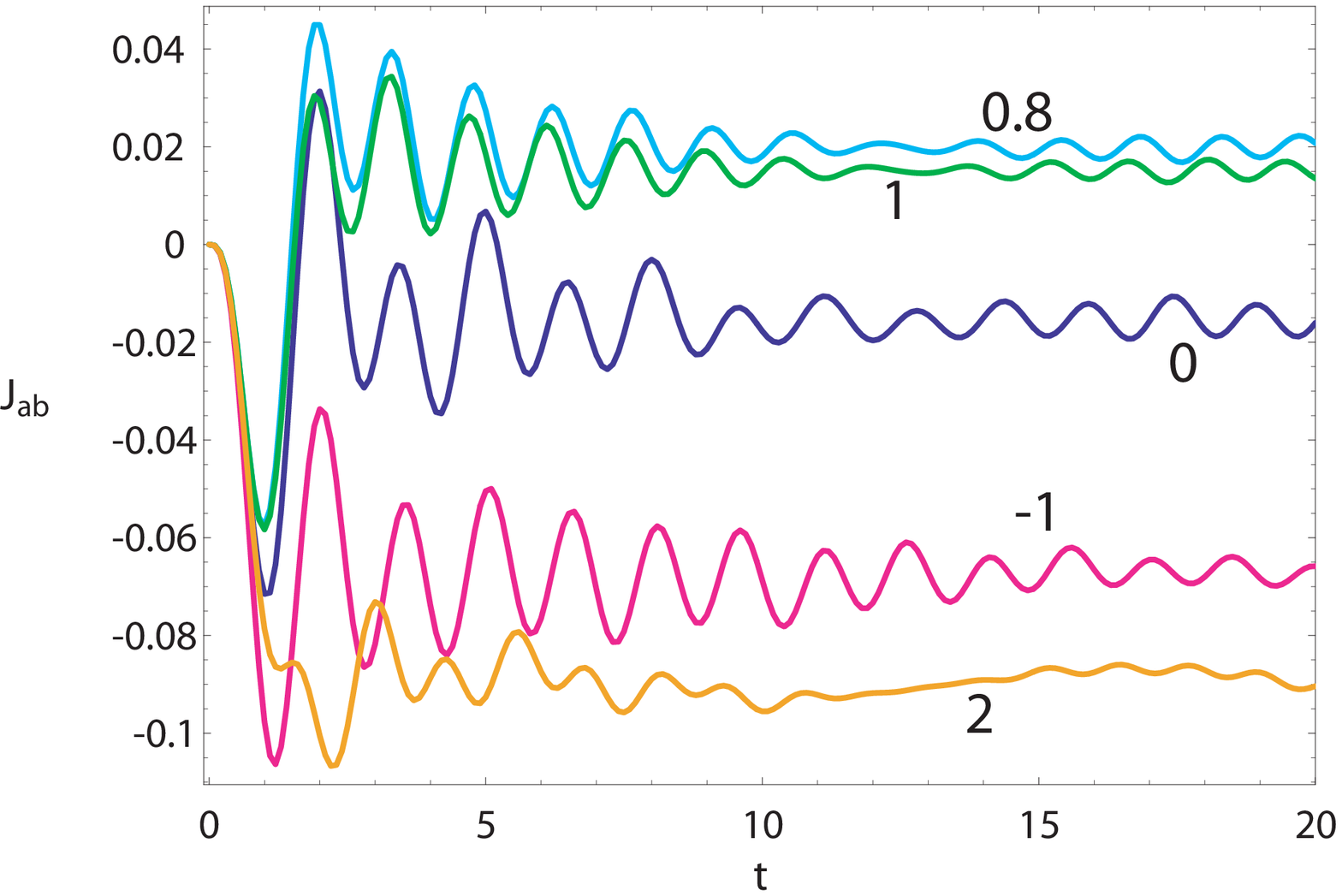}
\caption{Time-dependence of  $\langle J_{ab}\rangle$. The numbers
shown close to each curve indicate  the values of $\epsilon_{c}$.
From bottom, $\e_{c}=2,-1,0,1,0.8$ in units of $t_{h}$.  } \label{vort1}
\end{figure}

\begin{figure}[htbp]
\includegraphics*[width=.47\textwidth]{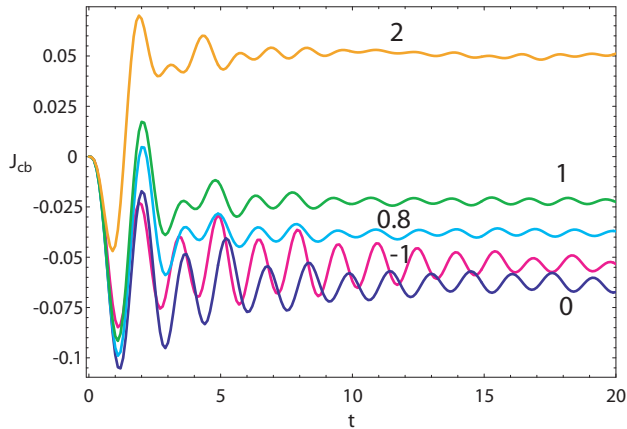}
\caption{Time-dependence of  $\langle J_{cb}\rangle$. The numbers
shown close to each curve indicate the values of $\epsilon_{c}$.
From bottom, $\e_{c}=0,-1,0.8,1,2$ in units of $t_{h}$. } \label{vort2}
\end{figure}

\subsection{Laminar flow and Vortices}
\label{vortices}

In Figures \ref{vort1} and \ref{vort2} we report the time-dependent
$\langle J_{ab}\rangle$ and  $\langle J_{bc}\rangle$   at times
$0<t<\frac{20\hbar}{t_{h}}$   for $\epsilon_a =\epsilon_{b}=0$ and
various  values of $\epsilon_c$. When the site $c$ is attractive for
electrons (that is, $\epsilon_c<0 $), despite some transient
oscillations, the current  remains negative, both on the $a$-$b$ and
$c$-$b$ bonds; the current flows from $b$ towards $a$ and $c$, so
the flow is {\em laminar}. The magnitude of the current on the
$a$-$b$ and $c$-$b$ bonds is comparable. Although site $c$ may have
a high  electron population, the local current does not concentrate
on either bond.

For    $\epsilon_c=0.8$  and  $\epsilon_c=1.0$  the current $J_{ab}$
on the $a$-$b$ bond (Figure 3), after a negative transient spike,
produces a positive one, which is the main contribution to the
bouncing current noted above. This behavior can be understood in
terms of a strong charge build-up on the $a$-$c$-$b$ arm of the ring
during the first burst following the switching of the bias, which
eventually triggers the temporary  back-flow. After the burst,
$J_{ab}$ remains positive  ($\e_{c}=0.8$ and  $\e_{c}=1.0$
curves in Figure 3) while
 $J_{cb}$  remains negative
value ($\e_{c}=0.8$ and  $\e_{c}=1.0$ curves in Figure 4). In other
terms, we observe the formation of an anti-clock-wise current {\em
vortex}.

Remarkably and unexpectedly, by increasing $\epsilon_{c}$
one reaches a critical value beyond which the vortex becomes
clock-wise, as one can see from the  $\e_{c}=2$ curves in Figures 3
and 4. This conclusion is unavoidable since the current changes sign
in both arms of the circuit. The total current (Figure 2) remains
negative, as one expects.  We were unable to find any simple
qualitative explanation for the inversion of the vortex.
In general the critical value of $\e_{c}$ depends on the bias $V$ and
for $V=0.5$ is about 1. Moreover, we emphasize that  the current across a
strongly repulsive site with $\epsilon_{c}=2$  is still comparable
in magnitude with the one on the $a$-$b$ bond.

\section{Many-body symmetry  and  Magnetic Response}\label{doubleminimum}

\begin{figure}[htbp]
\includegraphics*[width=.47\textwidth]{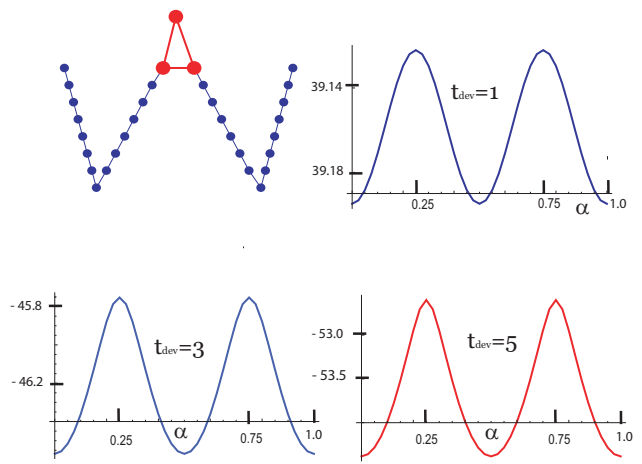}
\caption{The ${\phi\over 2}$ periodicity arising from the
symmetry-induced false pairing. Top left: the triangular device
connected to  $\L=$14-site wires. The other panels  show the ground
state energy $E_{0}(\a)$
 versus $\alpha.$ Top right: $t_{dev}=t_{h}$. Bottom left:
$t_{dev}=3 t_{h}$. Bottom right: $t_{dev}=5 t_{h}$. } \label{bicorni}
\end{figure}

Nanoring devices asymmetrically connected to wires of  $\L$ sites
each are even more peculiar for their magnetic properties.  Here we
restrict to the case $\e_{c}=0$. The magnetic flux
$\phi$ through the ring is inserted by the Peierls prescription\cite{topics},\cite{canright}
\begin{equation}
t_{ac}\rightarrow t_{ac} e^{2\pi i\a},
\end{equation}
where $\a=\frac{\phi}{\phi_{0}}$ and  $\phi_{0}=\frac{h c}{e}$ is
the magnetic flux quantum.  In the case of vanishing bias, in
stationary conditions one finds a diamagnetic current
which is confined to the triangular ring.
 In Figure  5  we illustrate our  results. We sketch a triangular device connected
to 14-site leads and the ground state energy $E_{0}(\a)$ versus
$\a=\frac{\phi}{\phi_{0}}$ for $t_{dev}\equiv
t_{ab}=t_{bc}=t_{ca}=1,3$ and 5.
 The expected
periodicity  in $\a$ with period 1 due to gauge invariance is of
course observed. Strikingly, however, we notice that the period is
actually ${1\over 2}$, and the system is diamagnetic, i.e. the
energy increases when the un-magnetized system is put in a magnetic
field. In other terms, $E_{0}(\a)$ looks like the ground state
energy of  a superconducting ring, although this is a non-interacting
model and the spectrum is gapless.

One could wander how the positive diamagnetic response at small fields
arises, since the initial dependence of the total energy on the flux
is quadratic and the second-order correction is always negative.
However this is an apparent paradox. The energy change is due to a
perturbation $t_{dev}(e^{2\pi i\a}-1)\sim t_{dev}(2\p
i\a-2\p^{2}\a^{2})$ which is the effect of changing the phase of one
bond. First order perturbation theory corresponds to compute the
current operator over the ground state and it is zero. The quadratic
contribution is given by second-order perturbation theory in
$2\p i\a t_{dev}$ plus a term coming from the first-order
correction in $-2\p^{2}\a^{2} t_{dev}$. The former term is always
negative while the second term can be either positive or negative.

\subsection{Nontrivial role of the wires}

The results in Figure \ref{bicorni} are striking because  the
isolated nano-ring with 3 sites at  half filling   does not simulate any superconducting behavior; instead, it  yields a  paramagnetic pattern, symmetric around $\alpha=\frac{1}{2}$  ($\alpha=1$ is equivalent to $\alpha=0$).
 One can easily work out   the lowest  energy
eigenvalue $E_{0}$   with 3 electrons.
  $E_{0}=-3$ for $\a=0$ sinks to
 $E_{0}=-2\sqrt{3}\sim -3.46$ at $\a={1\over 4},$ and raises
again to $E_{0}=-3$ at $\a={1\over 2}.$ Thus, there is a half-fluxon
periodicity at half filling, but $\a=0$ and $\a={1\over 2}$ are
maxima and correspond to degenerate 3-body states.
\begin{figure}[htbp]
\includegraphics*[width=.27\textwidth]{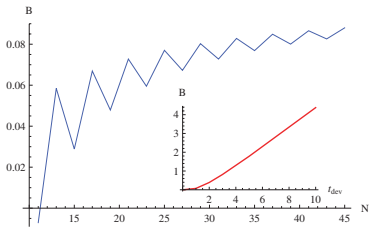}
\caption{Dependence of the barrier  hight  $B$ on the total number
{\cal N}
of sites, which include the triangular device and the $L$ and $R$
leads.  The parameters are the same as in Fig. 6. The
line is a guide for the eye. Note that for short leads $B$  is negative. The inset shows the almost linear
dependence of $B$ on $t_{dev}=t_{ab}=t_{ac}=t_{bc}$.} \label{HvsN}
\end{figure}

The  superconductor-like response  requires the presence of wires,
despite the  fact that the diamagnetic currents induced by the field
are strictly confined to the triangular device. The currents do not visit the
wires, but the electron wave functions do. In order to produce the
double minimum, the wires must be rather long.
The barrier height   $B=E_{0}(\frac{1}{4})-E_{0}(0)$
depends on the total number of atoms
${\cal N}=3+2\L$ (see  Figure 6) and below a minimum length $B<0$.
In Fig. \ref{HvsN} we plot the dependence of the barrier $B$ on the
number of sites ${\cal N}=3+2\L$ of the leads. We observe that $B$ saturates with
increasing ${\cal N}$.
This finding is noteworthy and unusual. The inset of Figure 6 shows how $B$ depends
on $|t_{dev}|$ and suggests that except for an initial quadratic
region the dependence is basically linear.
 With hopping
integrals in the eV range $B$   easily exceeds room temperature.

\subsection{Bipartite and not bipartite wired devices}
Next we investigate  why the effect takes place in this geometry and
at half filling.  A crucial observation is that the system depicted
in Figure 5 is not a bipartite graph. By contrast, Figure 7 shows
the flux dependence of the ground state of a bipartite graph,
namely, a square device connected to wires. The response is
paramagnetic since the  ground state at $\phi=0$ is degenerate, the
field lifts the degeneracy, a Zeeman effect occurs, and the ground
state energy is lowered. In addition, the trivial periodicity is
observed.
\begin{figure}[htbp]
\includegraphics*[width=.33\textwidth]{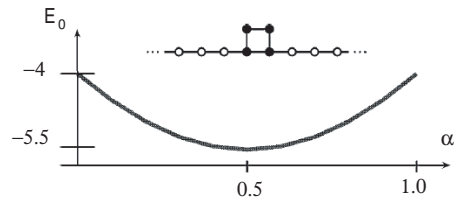}
\vspace{.5cm} \caption{Flux-dependent energy $E_{0}(\a)$ for a
square
  device with  $t_{dev}=t_{h}$.
   Notice the  scale, the {\em paramagnetic} behavior, and the level crossing at no flux.
  The central  part of the system, with the device and the beginning of the wires, is also shown.}
   \label{quadrato}
   \end{figure}
In Figure 8  we add one site to the device, and   the diamagnetic
double minimum pattern is found for  the resulting  pentagonal
device, which does not produce a bipartite graph.
Based on this observation and on a symmetry analysis, we can show that the half-fluxon periodicity ($E_{0}(\frac{1}{2})=E_{0}(0)$)  holds. This is a theorem which holds for any ring with an odd number of atoms connected to leads of any length provided that the site energies vanish.\\
\begin{figure}[htbp]
\includegraphics*[width=.27\textwidth]{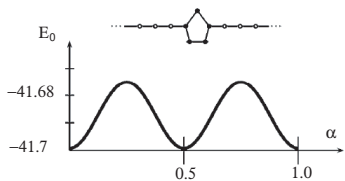}
\caption{Flux-dependent energy $E_{0}(\a)$ for a pentagonal device
with  $t_{dev}=t_{h}$.
   Notice  the {\em diamagnetic} behavior, and the  absence of level crossing at no flux.
  The central  part of the system, with the device and the beginning of the wires, is also shown.} \label{HvsN}
\end{figure}\label{pentagono}

\subsection{Symmetry analysis}

Let  $C$ denote the charge conjugation operation (or electron-hole
canonical transformation) $c_{\a}\rightarrow b_{\a}^{\dagger}$,
where $c_{\a}$ annihilates electrons and $b^{\dagger}_{\a}$ creates
a hole with quantum numbers $\a$.   $C$ is  equivalent to
     $\t_{mn}\rightarrow -\t_{mn}$ throughout, or, since the site energies vanish, to $H\rightarrow
     -H$.

 We recall  two elementary  results of graph theory:  1)  an isolated  ring   is  a bipartite graph if and only if  it has  an even number of atoms 2)  adding  wires of any length does not change the result.

      In bipartite lattices, $C$ is equivalent to a sign change of alternating orbitals,
      which is a gauge transformation. Hence, $H$ and
     $-H$ have the same one-body spectrum, that is, the spectrum is top-down symmetric.

       For the wired triangular ring and any other non-bipartite graph,
        the one-body spectrum is {\em not } top-down symmetric (except\cite{except}
        at $\a=\frac{1}{4},\a=\frac{3}{4}$)
 so it does not appear the  same after the transformation, and $C: H\rightarrow -H $
 is equivalent to changing the sign of one bond in the ring.  But
 this is just the effect of the operation $F:  \a \rightarrow \a+\frac{1}{2}$ which inserts half a fluxon in
the ring. Therefore  the combined operation $C\circ F$ is an exact
symmetry of the many-electron state  which holds at half filling.
It is clear that   $F: H\rightarrow -H $   i.e. $F$ turns the spectrum upside-down.

However, as
noted above, the spectrum is {\em not } top-down symmetric and when
at $\phi={\phi_{0}\over 2}$ the occupied and empty spin-orbitals
have exchanged places the system does not quite look like it was at
$\a=0.$  For instance, at the top of the spectrum of  the  device of Figure 5     at $ \a=0$ there is an empty split-off state.  At $ \a=0.5$ this  becomes a deep  state below the band.
Nevertheless, we wish to prove that  the many-body ground state energy $E_{0}(\a)$ at
$\a={1\over 2}$ is exactly the same as at $\a=0.$

 The reason lies in
another symmetry of the many-body state  at half filling.
In terms of one-body spin-orbital levels, Tr$ H(\a)=0$ implies \be
E_{0}(\a)=\sum_{i}n_{i}(\a)\e_{i}(\a)=- \sum_{i}(1-n_{i}(\a))\e_{i}(\a).\ee
Since under $F$ the negative of the energy of the unoccupied
levels coincide with the energy of the occupied ones, we have
$\e_{i}(\frac{1}{2})=-\e_{i}(0)$ and
$(1-n_{i}(\frac{1}{2}))=n_{i}(0)$ from which it follows that
  \be E_{0}(\frac{1}{2})= -
\sum_{i}(1-n_{i})\e_{i}(\frac{1}{2})=E_{0}(0) .\ee

\begin{figure}[htbp]
\includegraphics*[width=.27\textwidth]{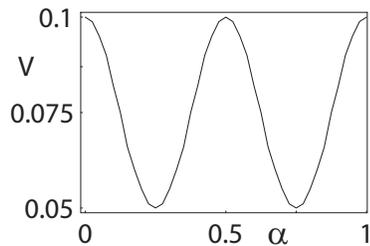}
\caption{Operation of the triangular device with
$\e_{a}=\e_{b}=\e_{c}=0$ and $t_{dev}=t_{h}$ used a a SQUID thread
by a flux $\phi$. A fixed current $J=0.016 \frac{t_{h}}{\hbar}$
flows through the device. The right wire site energies are raised by
$V_{R}=V(\phi)$, the left wire site energies are lowered by
$V_{L}=-V(\phi)$, and $V$ is adjusted in order to keep $J$ fixed.
The  $V(\phi)$ versus $\phi$ plot is perfectly periodic with a
half-fluxon period and  simulates a superconducting SQUID, although
no superconductors are needed.} \label{squidtriangolo}
\end{figure}
\begin{figure}[htbp]
\includegraphics*[width=.27\textwidth]{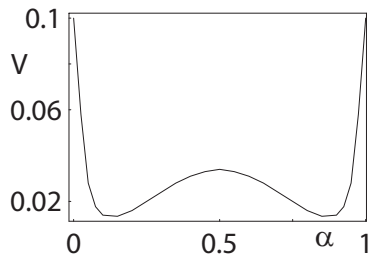}
\caption{ Here we present the same thought experiment as in the
previous Figure, except that the device is a square with
$t_{dev}=t_{h}$ threaded by a flux $\phi$ with zero on-site
energies.
 The SQUID-like behavior and the half-fluxon periodicity are lost,
 as expected in bipartite graphs (see text).
  In this particular example the lower-left bond was raised to
   $1.2 t_{h}$ in order to study the effects of a slight
   deformation of the square, but no relevant departure from the
   perfect square behavior was found.}\label{nonsquid}
\end{figure}
\subsection{Superconductor-free Quantum Interference Device}\label{squid}
The superconducting quantum interference
device (SQUID) consists of two superconductors separated by thin
insulating layers to form two parallel Josephson junctions, and
can be used as a magnetometer to detect tiny magnetic fields. If a
constant biasing current is maintained in the SQUID device, the
measured voltage oscillates with the  change in the magnetic flux,
and by counting the oscillations one can  evaluate the flux change
which has occurred. In principle  one can produce
Josephson-like oscillations without the need of superconductivity;
if the device can be realized this can be a physical principle  of
interest for applications, maybe even  at room temperature. Impurities and any disturbance affecting the quantum coherence can lower the operating temperature, but since in principle $B$ can be as large as 1eV we believe there is enough motivation for an experimental activity on this idea.
According to the above arguments, the triangular pentagonal and
other odd-numbered polygonal devices should  simulate a SQUID,
while bipartite graphs should present a normal behavior. We have
performed the thought experiment with the triangular device
connected to infinite wires; the results are reported in Figure
\ref{squidtriangolo}.
 We keep a fixed current $J$  flowing  through the device by adjusting $V(\phi)$,  while the flux $\phi$ threading the device is varied. It can be seen that  the  plot of  $V(\phi)$ versus $\phi$ is periodic with a half-fluxon period. The system   simulates a SQUID, although  no superconductors are needed.   A counter-example is given in Figure \ref{nonsquid}, where the triangular device is replaced by a square one and the effect disappears.
\section{Conclusions}\label{conclusions}
We have shown that simple asymmetric closed circuits have rather subtle  properties when unsymmetrically connected to the biased circuit, that could also be useful for designing new kinds of  devices.
We have found that  an impurity site of energy $\varepsilon_{c}$ in the longer arm of a triangular device can cause a transient bouncing current, which goes in the opposite direction than the long-time current  and is much more intense but lasts for a short time.  Looking at the current distribution inside the device, $\varepsilon_{c} <0$ favors a laminar current flow; instead,  $\varepsilon_{c} >0$ produces an anticlockwise current vortex; when a critical value is exceeded, however, the vortex
 chirality reverses.
Further, we have shown that   a class of  closed circuits at half filling have a degenerate ground state
which is paramagnetic, i.e. gains energy in a magnetic field by a
Zeeman splitting.  The magnetic
 behavior is completely changed when the circuits are connected to leads and form a non-bipartite lattice.
  Long leads  constitute an essential requirement and although the diamagnetic currents are confined to the ring, the leads
    modify the magnetic properties substantially. With long enough wires one obtains a diamagnetic behavior
     with half fluxon periodicity and a robust barrier separating the energy  minima at 0 and
     ${\phi_{0}\over 2}$.  This pattern mimics a superconducting ring although there are no gap,
     no interactions, no critical temperature. We have traced back the origin  of this fake superconducting behavior
     to a combination  of charge-conjugation and flux which provides a symmetry of  the many-electron determinantal state. Finally we have shown
      that in principle one can extend  this simulation to the point of building a functioning interference
       device capable of measuring local fields and analogous to a
       SQUID but working even above
       room temperature. It cannot be excluded that even-sided
       circuits can be useful for the same purpose, and indeed
       Figure 10 suggests that one could do so, exploiting the
       trivial periodicity. The signal in Figure 9, however, is much
       more monochromatic, and this suggests that the odd-sided
       version should make it much easier to read-off the magnetic
       field intensity from the amplitude of the voltage
       oscillation.

\vspace{3cm}


\begin{thebibliography}{10}
\bibitem{except}As noted above, the one-body spectrum is not generally top-down
     symmetric; at $\a=0$, out of the $2\L+3$ eigenvalues, $\L+2$ are
     negative and $\L+1$  positive; at $\a=0.5$, because of the
     turnover of the spectrum,  $\L+2$ are positive
      and L+1 negative. At $\a=0.25$, by a gauge transformation on $H$ one can change the signs of all real bonds, then by complex conjugation one obtains $-H$; since $H$ and $H^{*}$ must have the  same (real) eigenvalues,  it follows that one eigenvalue vanishes and
      the spectrum is  top-bottom symmetric like in bipartite
      lattices.
\end{thebibliography}
\end{document}